\documentclass[prd,aps,preprint,amsmath,amssymb,eshowkeys,nofootinbib]{revtex4-2}
\usepackage[mathscr]{euscript}
\usepackage{dcolumn}
\usepackage{bm}
\usepackage{graphicx}
\usepackage{subfigure}
\usepackage{changepage}
\usepackage{tabularx}
\usepackage{amsmath,amssymb,amsthm}
\usepackage[colorlinks=true,linkcolor=red,urlcolor=green,citecolor=green]{hyperref}
\newcommand{\bea}{\begin{eqnarray}}
\newcommand{\eea}{\end{eqnarray}}
\newcommand{\beq}{\begin{equation}}
\newcommand{\eeq}{\end{equation}}

\def\/{\over}

\begin{document}

\title{Holographic Inflation and Slow-Roll Inflation Within Rényi Entropic Framework in the Light of ACT DR6}
\author{Qihong Huang$^{1}$\footnote{Corresponding author: huangqihongzynu@163.com}, He Huang$^{2}$, Hao Chen$^{1}$, and Qingdong Wu$^{1}$}
\affiliation{
$^1$ School of Physics and Electronic Science, Zunyi Normal University, Zunyi, Guizhou 563006, China\\
$^2$ College of Mechanical and Electrical Engineering, Jiaxing Nanhu University, Jiaxing, Zhejiang 314001, China
}

\begin{abstract}
Based on the Rényi entropy, Rényi holographic dark energy has been proposed to explain the current accelerated expansion of the universe. In this paper, we analyze holographic inflation and slow-roll inflation within the framework of Rényi holographic dark energy (RHDE) using ACT DR6. Our results show that holographic inflation is ruled out by the data, while slow-roll inflation with power-law potentials for $n=\frac{1}{2}$ and $n=\frac{1}{3}$ is viable for a suitable choice of $N$ and $C$. We also analyze the inflationary attractor and confirm its existence. In addition, we compute the primordial power spectrum and find it falls well within the observational bounds. Thus, slow-roll inflation is favored in RHDE, but holographic inflation is not.
\end{abstract}

\maketitle

\section{Introduction}

Inflation, an exponentially accelerating expansion in the early universe, successfully resolves the horizon and flatness problems encountered in the standard big bang \mbox{model~\cite{Guth1981, Linde1982}}. Inflation creates adiabatic, Gaussian, and nearly scale-invariant scalar perturbations. These perturbations then seed the Cosmic Microwave Background (CMB) temperature anisotropies and the formation of Large-Scale Structure~\cite{Mukhanov1981, Lewis2000, Bernardeau2002}. This inflationary picture is strongly supported by observational results from COBE~\cite{Smoot1992}, WMAP~\cite{Hinshaw2013}, Planck~\cite{Planck2020}, and ACT DR6~\cite{Louis2025}. The simplest inflation model, called slow-roll inflation~\cite{Noh2001, Weinberg2008}, features a canonical scalar field slowly rolling down its potential to drive exponential expansion. After inflation, the primordial scalar power spectrum encodes all information about scalar perturbations, with the scalar spectral index $n_{s}$ describing its scale dependence~\cite{Planck2020a}. Meanwhile, tensor perturbations generated during this epoch give rise to primordial gravitational waves, whose amplitude is characterized by the tensor-to-scalar ratio $r$~\cite{Planck2020a, Maggiore2018}. With these two measurable quantities, $n_{s}$ and $r$, one can tightly constrain slow-roll inflation models using the current observational data~\cite{Planck2020a}. Under this observational framework, numerous slow-roll inflation models~\cite{MYuennan2026, Ellis2026, Ketov2025, Keskin2025, Huang2025, Ding2024, Ragavendra2024, Pozdeeva2024, Zhang2024, Marco2024, Lambiase2023, Afshar2023, Bhat2023, Dioguardi2022, Karciauskas2022, Chen2022, Capozziello2021, Forconi2021, Cai2021, Gamonal2021, Fu2020, Akin2020, Fu2019, Granda2019, Gonzalez-Espinoza2019, Granda2019a, Yi2018, Casadio2018, Odintsov2018, Tahmasebzadeh2016, Yang2015, Koh2014, Gao2014, Antusch2014, Guo2010, Satoh2010, Kaneda2010, Brax2009, Tzirakis2009, Barenboim2007, Peiris2006, Gong1999} and potentials~\cite{Yi2025, Choudhury2024, Choi2022, Fei2020, Barrow2018, Gao2018, Koh2017, Fei2017, Lin2016, Cline2006, Easther2003, Adams1995, Liddle1994} have subsequently been developed. Beyond the slow-roll paradigm, other models including constant-roll inflation~\cite{Motohashi2015, Gao2017, Gao2018a, Yi2018a, Gao2018b, Motohashi2019, Gao2019, Ravanpak2022, Mohammadi2022, Shokri2022, Panda2023, Liu2024, Huang2025a}, curved inflation~\cite{Thavanesan2021, Shumaylov2022, Huang2022, Huang2023a}, and slow expansion~\cite{Liu2011, Liu2013, Cai2016} have also been constructed. These alternatives can generate nearly scale-invariant primordial power spectra in agreement with observations.

{Holographic dark energy (HDE) has emerged as a compelling candidate to explain the nature of dark energy~\cite{Cohen1999}. This scenario originates from applying the holographic principle~\cite{Susskind1995, Bousso2002}, a profound conjecture rooted in black hole thermodynamics and quantum gravity, to the cosmological framework. However, the original HDE model does not produce a viable late-time acceleration and thus fails to describe the current expansion history of the universe~\cite{Hsu2004, Li2004}. To rectify this situation, several new HDE models have been proposed, establishing the HDE framework as a physically viable dark energy candidate that has attracted extensive study. For the HDE, horizon entropy provides the foundation, with different choices of horizon entropy giving rise to distinct HDE models~\cite{Wang2017}.} Based on the Rényi entropy, Rényi holographic dark energy (RHDE) has been proposed~\cite{Moradpour2018}. Compared to the original HDE model~\cite{Hsu2004}, this model has succeeded in explaining the current accelerated phase of the universe, and has received extensive attention in both theoretical~\cite{Ghaffari2019, Iqbal2019, Sharma2020, Sharma2020a, Saha2021, Shekh2021, Sardar2021, Saha2023, Manoharan2023, Das2025, Abdelrashied2025} and observational studies~\cite{Prasanthi2021}.

Recently, holographic dark energy (HDE) drives an inflationary model called holographic inflation~\cite{Nojiri2019}, which also generates nearly scale-invariant primordial power spectra and receives support from observations. {In slow-roll inflation, the near exponential expansion is driven by a canonical scalar field that rolls slowly down a nearly flat potential. In contrast, holographic inflation does not introduce any scalar field or an explicit potential. Instead, the accelerated expansion is driven by HDE, whose specific energy density evolution replaces the role of the scalar potential. The resulting slow-roll parameters are controlled by the IR cutoff scale rather than by the shape of a potential.} In the original holographic inflation model, the particle horizon and the future event horizon were adopted as the IR cutoff. Later studies moved beyond the particle or future event horizon by adopting the {Granda and Oliveros (GO) length scale}~\cite{Granda2008, Granda2009} as the IR cutoff. Within this IR cutoff, holographic inflation proves viable for both HDE~\cite{Oliveros2019} and Tsallis HDE~\cite{Mohammadi2021}. In contrast, in the framework of anti-de Sitter black hole entropy with the Hubble horizon as the IR cutoff, a specific HDE model is constructed, but holographic inflation within this model cannot be supported by observations~\cite{Huang2025}. However, when slow-roll inflation is studied in this framework, the power-law potential can be supported by observations~\cite{Huang2025}, whereas in general relativity it is excluded by Planck 2018 results~\cite{Planck2020a} and disfavored by ACT DR6~\cite{Calabrese2025}.

Under the Rényi cosmology framework, which was previously used to describe late-time acceleration, the vacuum case yields a de Sitter inflationary expansion without requiring an inflaton field~\cite{Ghaffari2019}. Within the RHDE framework, questions arise as to whether holographic inflation is supported by observations and whether slow-roll inflation with a power-law potential is supported by observations. This paper will address these issues. This paper has two main objectives: to investigate holographic inflation and to examine slow-roll inflation in the RHDE model. The paper is organized as follows: In Section II, we examine holographic inflation in the RHDE framework. In Section III, we analyze slow-roll inflation in the same model. Finally, our main conclusions are presented in Section IV.

\section{Holographic Inflation}

By employing the Rényi entropy, the energy density of RHDE is given as~\cite{Moradpour2018}
\beq
\rho_{de}=\frac{3C^{2}H^{2}}{\kappa^{2}\big(1+\frac{8\pi^{2}\delta}{\kappa^{2} H^{2}}\big)},\label{rhode}
\eeq
with $\kappa^2=8\pi G$, $C$ being a dimensionless constant and $\delta$ being the model parameter. Here, under $\kappa^{2}=1$ with $c=1$, $\delta$ has the same dimension as $H^{2}$. To explain the current accelerated phase of the universe, $\delta$ takes a value in the range from $-1400$ to $-900$~\cite{Moradpour2018}. For $\delta=0$, the energy density of RHDE reduces to that of the original HDE~\cite{Hsu2004}.

We consider a spatially flat, homogeneous, and isotropic Friedmann--Robertson--Walker universe described by the metric
\beq
ds^{2}=-dt^{2}+a^{2}(t)(dr^{2}+r^{2}d\Omega^{2}),
\eeq
with $a(t)$ denoting the scale factor and $t$ the cosmic time. The Friedmann equations can be expressed as
\bea
& H^{2}=\displaystyle\frac{\kappa^{2}}{3}\big(\rho_{\phi}+\rho_{de}\big),\label{H2}\\
& 2\dot{H}+3H^{2}=-\kappa^{2}(p_{\phi}+p_{de})\label{H1},
\eea
with $\rho_{\phi}=\frac{1}{2}\dot{\phi}^{2}+V(\phi)$ and $p_{\phi}=\frac{1}{2}\dot{\phi}^{2}-V(\phi)$ being the energy density and pressure of a scalar field that obeys
\beq
\ddot{\phi}+3H\dot{\phi}+V_{,\phi}=0.\label{phi2}
\eeq
Similarly, $\rho_{de}$ and $p_{de}$ are the energy density and pressure of RHDE, which satisfy the conservation equation
\beq
\dot{\rho}_{de}+3H(\rho_{de}+p_{de})=0,\label{cde}
\eeq
where $p_{de}=\omega_{de}\rho_{de}$, with $\omega_{de}$ being the equation of state parameter.

To investigate holographic inflation in the framework of RHDE, we consider inflation driven exclusively by HDE, with the scalar field $\phi$ being negligible. Thus, the Friedmann Equations~(\ref{H2}) and~(\ref{H1}) take the form
\bea
& H^{2}=\displaystyle\frac{\kappa^{2}}{3} \rho_{de},\\
& 2\dot{H}+3H^{2}=-\kappa^{2} p_{de},
\eea
from which $\dot{H}$ can be derived as
\beq
\dot{H}=-\frac{3}{2}H^{2} \Big( 1 + \frac{C^{2}}{1+\lambda}\omega_{de} \Big).\label{H11}
\eeq
with
\beq
\lambda=\frac{8\pi^{2}\delta}{\kappa^{2}H^{2}}.
\eeq

Combining Equation~(\ref{rhode}) with Equation~(\ref{cde}) leads to
\beq
\omega_{de}=-1-\frac{2}{3}\Big( 1+\frac{\kappa^{2} \lambda \rho_{de}}{3 C^{2}} \Big)\frac{\dot{H}}{H^{2}}.\label{ode}
\eeq
Substituting the relation from Equation~(\ref{ode}) into Equation~(\ref{H11}) yields
\beq
\dot{H}=-\frac{3}{2}H^{2} \Big[ \frac{1+\lambda+(1-C^{2})(1+\frac{1}{\lambda})}{1+\lambda+(1-C^{2})(1+\frac{1}{\lambda})-C^{2}} \Big].\label{H111}
\eeq

To investigate the inflation model, one can directly calculate the slow-roll parameters $\epsilon_n$, which are expressed in terms of the Hubble parameters $H$ and $\dot{H}$~\cite{Martin2014}. {Then, following Ref.~\cite{Nojiri2019, Oliveros2019, Mohammadi2021, Huang2025}, by means of Equation~(\ref{H111})}, the first and second slow-roll parameters can be derived and expressed as
\bea
& \epsilon_{1}=\displaystyle\frac{3}{2} \Big[ \frac{1+\lambda+(1-C^{2})(1+\frac{1}{\lambda})}{1+\lambda+(1-C^{2})(1+\frac{1}{\lambda})-C^{2}} \Big],\\
& \epsilon_{2}=\displaystyle\frac{3C^{2}\big[(1-C^{2})\frac{1}{\lambda}-\lambda \big]}{\big[ 1-C^{2}+\lambda+(1-C^{2})\big(1+\frac{1}{\lambda}\big) \big]^{2}}.
\eea
With these slow-roll parameters, the inflationary observables, namely the scalar spectral index $n_{s}$ and the tensor-to-scalar ratio $r$, are expressed as~\cite{Martin2014}
\bea
& n_{s}=1-2\epsilon_{1}-\epsilon_{2},\label{nsh}\\
& r=16\epsilon_{1}.\label{rh}
\eea

\begin{figure}[h]
%\begin{center}
\includegraphics[width=0.6\textwidth]{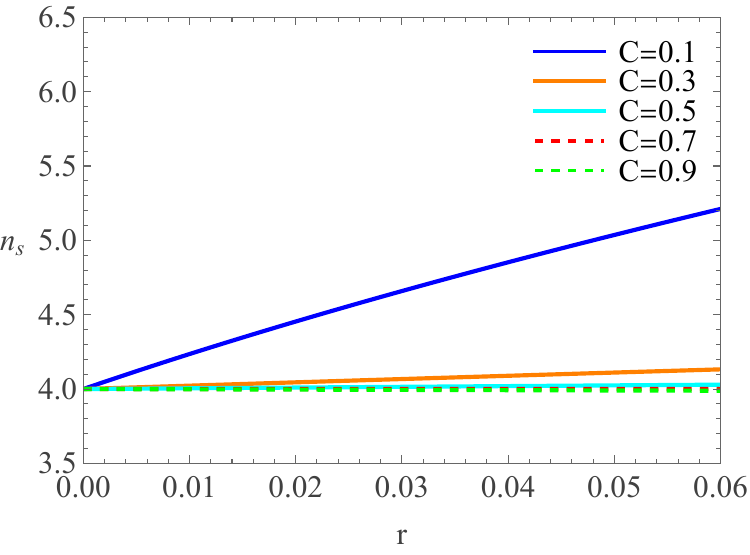}
\caption{\label{Fig1} Relation between $n_{s}$ and $r$ for holographic inflation within RHDE.}
%\end{center}
\end{figure}

Solving Equation~(\ref{rh}) yields the solution for $\lambda$; then, by substituting $\lambda$ into Equation~(\ref{nsh}), one can write $n_{s}$ as a function of $r$. The relation between $n_{s}$ and $r$ is plotted in Figure~\ref{Fig1}, with $C$ being the only free parameter. This figure shows that as $C$ increases, the lines overlap and become {indistinguishable for $C > 0.7$}. As can be seen in this figure, $n_{s}$ is {larger than $4.0$}, which is significantly higher than the Planck 2018 constraint $n_{s}=0.9668 \pm 0.0037$~\cite{Planck2020} and the ACT DR6 constraint $n_{s}=0.9743 \pm 0.0034$~\cite{Louis2025}, indicating that this model is ruled out by observations. Consequently, similar to the case in the HDE model based on anti-de Sitter black hole entropy~\cite{Huang2025}, within the framework of RHDE, holographic inflation is also not observationally viable. It should be noted, however, that holographic inflation is realized in other HDE models employing different IR cutoffs~\cite{Nojiri2019, Oliveros2019, Mohammadi2021}.

\section{Slow-Roll Inflation}

In the previous section, we have demonstrated that holographic inflation in the RHDE model is inconsistent with observations from Planck 2018 results and ACT DR6. In this section, we examine whether slow-roll inflation within RHDE is consistent with observational constraints. 

\subsection{Slow-Roll Dynamics}

{In the standard slow-roll inflation, under the slow-roll approximations $\frac{1}{2}\dot{\phi}^{2} \ll V(\phi)$ and $|\ddot{\phi}| \ll |3 H \dot{\phi}|$, one can obtain the relation $H^{2} \sim V(\phi)$, which indicates $V(\phi)$ is of the same order of magnitude as $H^{2}$. Then, in the framework of RHDE, under the slow-roll approximations, Equations~(\ref{H2}) and~(\ref{phi2}) can be expressed as
\bea
& H^{2} \approx \displaystyle\frac{\kappa^{2}}{3}V+\frac{C^{2}H^{2}}{1+\frac{8\pi^{2}\delta}{\kappa^{2}H^{2}}},\label{H20}\\
& 3H\dot{\phi}+V_{,\phi} \approx 0.\label{phi20}
\eea
From the above equation, and given that $\delta$ is much smaller than the inflationary scale $H^{2}$, we can conclude $\delta \ll H^{2}$ as a self-consistency condition. Under this condition, the dynamical equations for RHDE approximately reduce to those in the standard HDE. Therefore, the effect of $\delta$ is negligible, and the perturbation equations are the same as in the standard HDE. In the standard HDE, it is assumed that inflation is driven by the scalar field and that the HDE only affects the background dynamics without influencing the perturbations~\cite{Chen2007, Wang2017}. Under this assumption, the scalar spectral index $n_{s}$ and the tensor-to-scalar ratio $r$ are expressed as~\cite{Chen2007}
\bea
& n_{s} = 1-8\epsilon+2\eta+2\sigma,\label{nsnsns}\\
& r = 16\epsilon.\label{rrr}
\eea
with
\bea
\epsilon = -\frac{\dot{H}}{H^{2}}, \qquad \eta = -\frac{\ddot{\phi}}{H \dot{\phi}},\qquad \sigma = \displaystyle\frac{\kappa^{2} \dot{\phi}^{2}}{H^{2}}.
\eea
Based on the above discussion, in order to study slow-roll inflation within the framework of RHDE, we adopt the same scalar spectral index ~(\ref{nsnsns}) and the tensor-to-scalar ratio ~(\ref{rrr}).}

Then, differentiating Equation~(\ref{H20}) yields
\beq
\dot{H}=-\frac{\kappa^{2}}{4(1-C^{2})}\Bigg[ 1+\frac{48(1-2C^{2})\pi^{2}\delta+2\kappa^{4}V}{2\sqrt{576\pi^{4}\delta^{2}+48\kappa^{4}(1-2C^{2})\pi^{2}\delta V+\kappa^{8}V^{2}}} \Bigg] \Big(\frac{V_{,\phi}}{3H}\Big)^{2}.\label{H21}
\eeq
Using Equations (\ref{H20}), (\ref{phi20}) and (\ref{H21}), the slow-roll parameters $\epsilon$, $\eta$ and $\sigma$ can be expressed as
\bea
& \epsilon =\displaystyle\frac{\kappa^{2}}{4(1-C^{2})}\Bigg[ 1+\frac{48(1-2C^{2})\pi^{2}\delta+2\kappa^{4}V}{2\sqrt{576\pi^{4}\delta^{2}+48\kappa^{4}(1-2C^{2})\pi^{2}\delta V+\kappa^{8}V^{2}}} \Bigg] \Big(\frac{V_{,\phi}}{3H^{2}}\Big)^{2},\\
& \eta = -\epsilon+\displaystyle\frac{V_{,\phi\phi}}{3H^{2}},\\
& \sigma = \kappa^{2} \Big(\displaystyle\frac{V_{,\phi}}{3H^{2}}\Big)^{2}.
\eea

To compute the values of $n_{s}$ and $r$, we take the power-law potential $V(\phi)$ to be of the form
\beq
V=V_{0}\phi^{n},\label{Vp}
\eeq
which is known as the chaotic potential, with both $V_{0}$ and $n$ being positive parameters. Then, the slow-roll parameters $\epsilon$, $\eta$ and $\sigma$ become

\noindent
\bea
& \epsilon = \displaystyle\frac{\kappa^{2}}{4(1-C^{2})}\Bigg[ 1+\frac{48(1-2C^{2})\pi^{2}\big(\frac{\delta}{V_{0}}\big)+2\kappa^{4}\phi^{n}}{2\xi} \Bigg]\Bigg[ \frac{2\kappa^{2}(1-C^{2})n \phi^{n-1}}{\kappa^{4}\phi^{n}-24\pi^{2}\big(\frac{\delta}{V_{0}}\big)+\xi} \Bigg]^{2},\label{epsilon1}\\
& \eta = \displaystyle-\epsilon+\frac{2\kappa^{2}(1-C^{2})n(n-1) \phi^{n-2}}{\kappa^{4}\phi^{n}-24\pi^{2}\big(\frac{\delta}{V_{0}}\big)+\xi},\label{delta1}\\
& \sigma = \displaystyle\kappa^{2}\Bigg[ \frac{2\kappa^{2}(1-C^{2})n \phi^{n-1}}{\kappa^{4}\phi^{n}-24\pi^{2}\big(\frac{\delta}{V_{0}}\big)+\xi} \Bigg]^{2},\label{sigma1}
\eea
with
\beq
\xi=\sqrt{576\pi^{4}\Big(\frac{\delta}{V_{0}}\Big)^{2}+48\kappa^{4}(1-2C^{2})\pi^{2}\Big(\frac{\delta}{V_{0}}\Big)\phi^{n}+\kappa^{8}\phi^{2n}}
\eeq
Then, substituting Equations (\ref{epsilon1}), (\ref{delta1}), and (\ref{sigma1}) into (\ref{nsnsns}) and~(\ref{rrr}), the scalar spectral index $n_{s}$ and the tensor-to-scalar ratio $r$ can be calculated. {By considering $\delta \ll H^{2}$, we obtain $\delta \ll V$ and adopt $10^{-4}<\frac{\delta}{V_{0}}<10^{-2}$ for the following discussion, a range which ensures that the RHDE correction to the standard HDE is small but non-negligible, allowing us to study its observational implications for inflation. For HDE, without CMB data the parameter $C$ is constrained to $C>1$, while the inclusion of CMB data forces $C<1$~\cite{Li2025}. Moreover, for $\frac{\delta}{V_{0}} \ll 1$, RHDE approximates HDE. Since this paper uses CMB data (ACT DR6) to analyze inflation, the constraints that include CMB data are more applicable. For these reasons, we adopt $C<1$ in the following discussion.}

{To examine the e-folds number $N$ during inflation, we adopt the range $50 \leq N \leq 70$, from which $N$ can be expressed as}
\beq
N = \ln \Big( \frac{a_e}{a_i} \Big) = \int^{t_e}_{t_i} H dt = -\int^{\phi_{e}}_{\phi_i} \frac{3H^{2}}{V_{,\phi}}d \phi.\label{N01}
\eeq
where the subscript $_{e}$ and $_{i}$ denote the times when inflation ended and began, respectively. $\phi_{e}$ is the scalar field at the end of inflation and can be solved from the relation $\epsilon_{e} \simeq 1$. By inserting $\phi_{e}$ into Equation~(\ref{N01}), we obtain $\phi_i$ as a function of $N$, but owing to the power exponent $n$, this function does not admit an analytical form and must be evaluated numerically. Then, substituting $\phi_{i}$ into the calculated expressions for $n_{s}$ and $r$, we can express $n_{s}$ and $r$ in terms of $N$. Different from the holographic inflation in the previous section, the expressions for $n_{s}$, and $r$ in slow-roll inflation depend on {the parameters $C^{2}$, $\frac{\delta}{V_{0}}$, $n$, and $N$}, and cannot be expressed analytically. Accordingly, we perform numerical calculations {for the cases $n=\frac{1}{2}$ and $n=\frac{1}{3}$} and display their relationships using figures. 

\begin{figure}[h]
%\begin{center}
\includegraphics[width=0.45\textwidth]{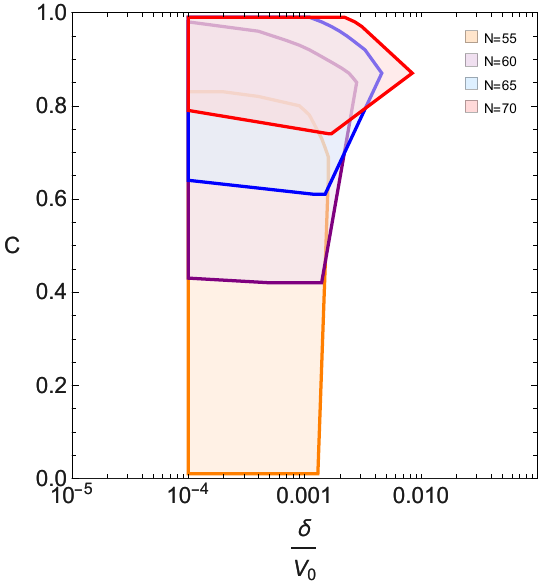}
\includegraphics[width=0.45\textwidth]{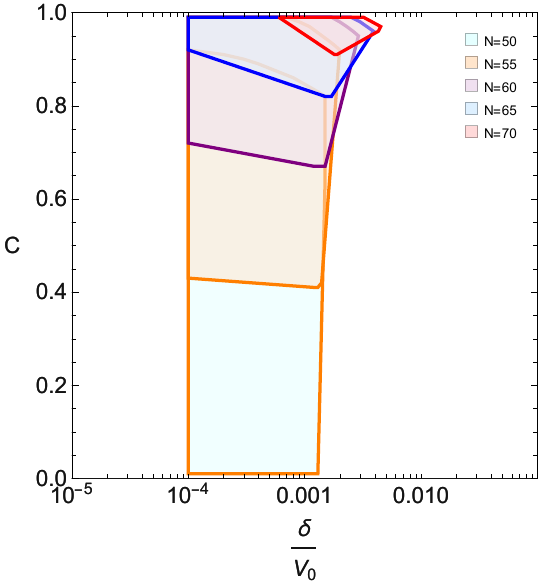}
\caption{\label{Fig2} Prediction regions in $C-\frac{\delta}{V_{0}}$ plane for slow-roll inflation within RHDE. The left panel is plotted for $n=\frac{1}{2}$, while the right one is for $n=\frac{1}{3}$.}
%\end{center}
\end{figure}

{To determine the values of the parameters and match our results with the observations, we adopt the constraints $r<0.038$~\cite{Calabrese2025} and $n_{s}=0.9743 \pm 0.0034$~\cite{Louis2025}. Then, for the cases $n=\frac{1}{2}$ and $n=\frac{1}{3}$, we plot the prediction regions in the $C-\frac{\delta}{V_{0}}$ plane with different e-fold numbers $N$, as shown in Figure~\ref{Fig2}; in the left panel, the case $n=\frac{1}{2}$ is plotted with $N=55,60,65,70$ since $N=50$ in the parameter space does not satisfy the constraints $r<0.038$ and $n_{s}=0.9743 \pm 0.0034$. These figures show that with the decrease in $N$, a larger $C$ is required, and the allowed region in the $C-\frac{\delta}{V_{0}}$ plane becomes smaller; as $n$ decreases, the allowed region for a given $N$ shifts upward. Specifically, for $n=\frac{1}{2}$ with $N=60$, $C>0.42$ is required; while for $n=\frac{1}{3}$ with $N=60$, $C>0.67$ is needed. Within the adopted range $10^{-4}<\frac{\delta}{V_{0}}<10^{-2}$, the results prefer a smaller value of $\frac{\delta}{V_{0}}$.}

\begin{figure}[h]
\begin{center}
\includegraphics[width=0.49\textwidth]{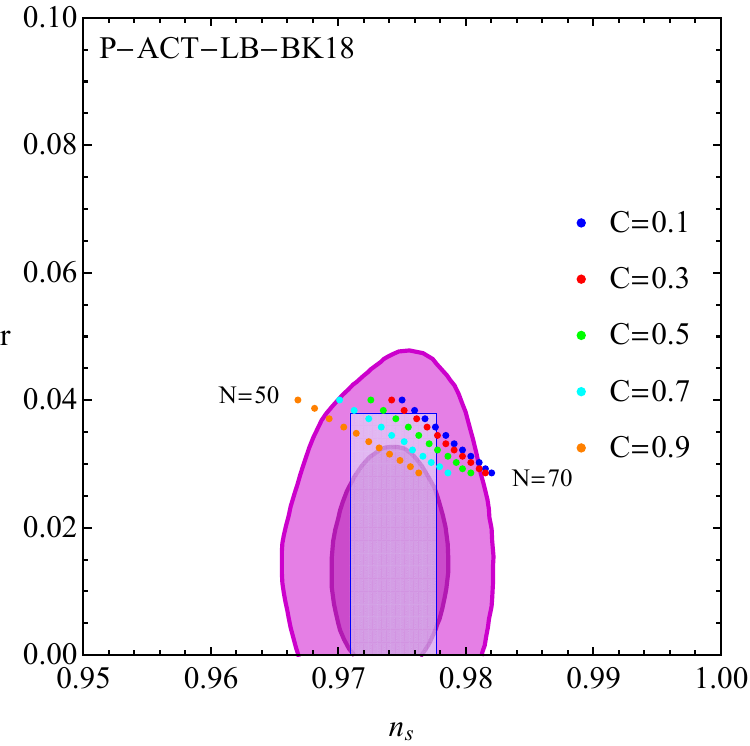}
\includegraphics[width=0.49\textwidth]{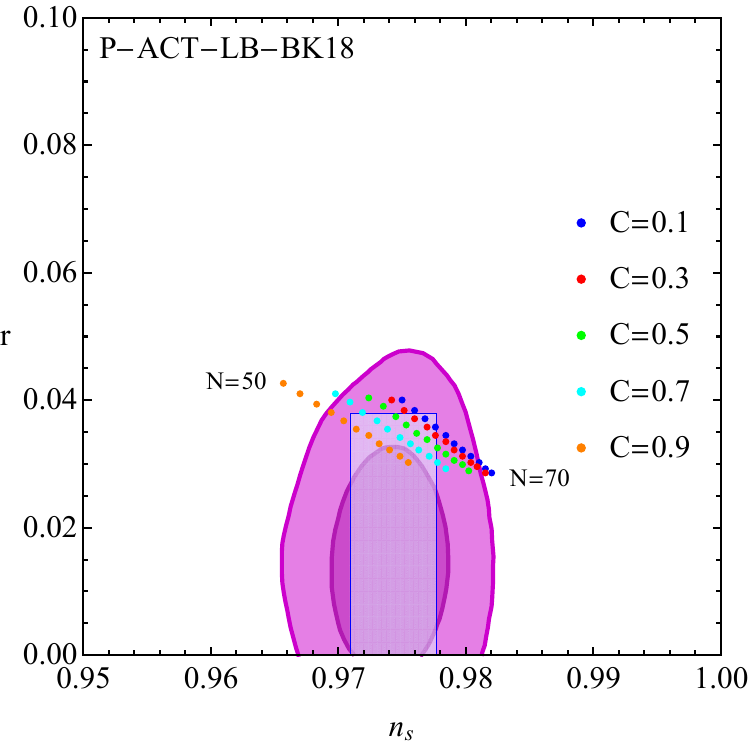}
\includegraphics[width=0.49\textwidth]{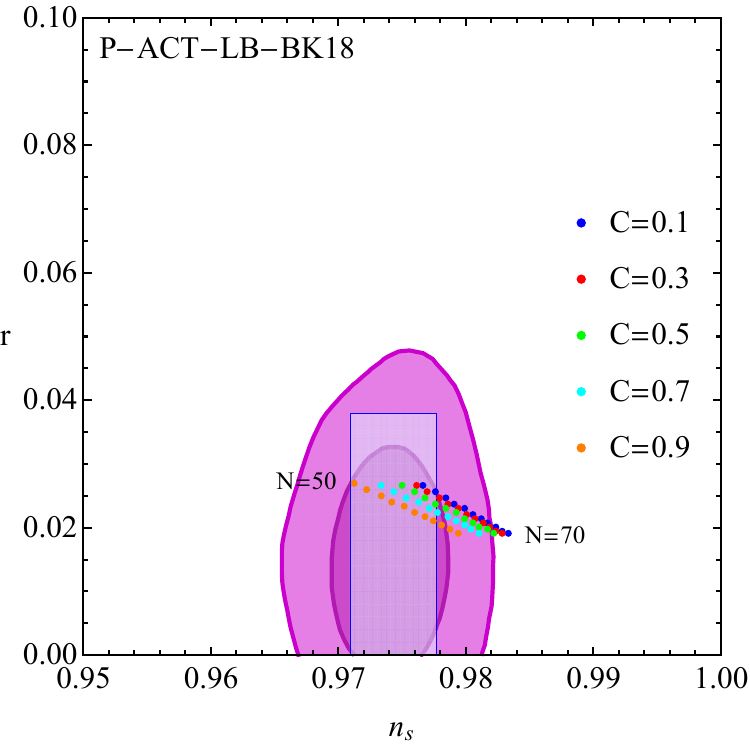}
\includegraphics[width=0.49\textwidth]{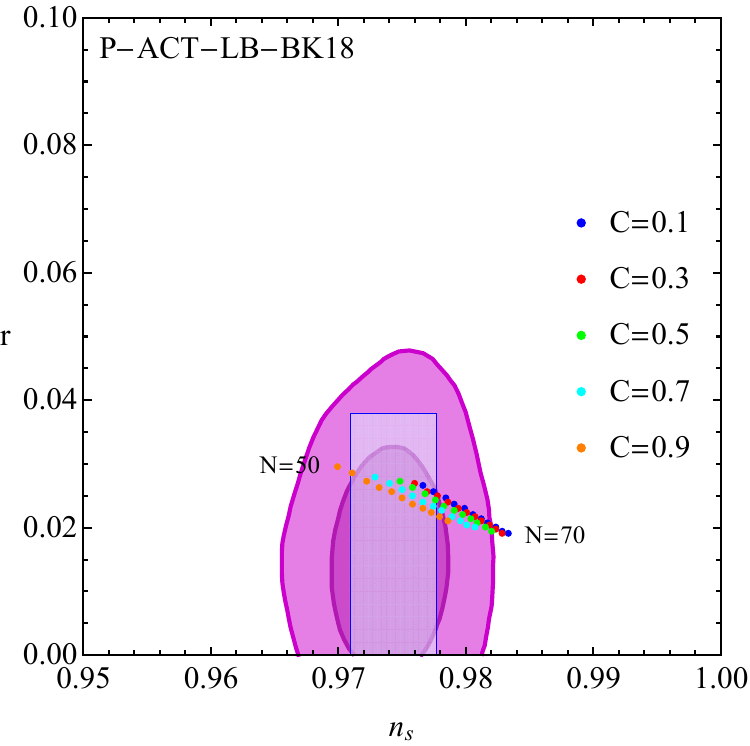}
\caption{\label{Fig3} Predictions power-law potential ~(\ref{Vp}) in the $r$-$n_{s}$ plane for four parameter sets: \linebreak  (1) $n=\frac{1}{2}, \frac{\delta}{V_{0}}=10^{-4}$, (2) $n=\frac{1}{2}, \frac{\delta}{V_{0}}=10^{-3}$, (3) $n=\frac{1}{3}, \frac{\delta}{V_{0}}=10^{-4}$, (4) $n=\frac{1}{3}, \frac{\delta}{V_{0}}=10^{-3}$.}
\end{center}
\end{figure}

{By fixing the parameters $\frac{\delta}{V_{0}}$ and $n$}, we plot the predictions of the power-law potential ~(\ref{Vp}) in the $r$-$n_{s}$ plane for four parameter sets in Figure~\ref{Fig2}, where we overlay our numerical results with the observational constraints from P-ACT-LB-BK18~\cite{Calabrese2025}, {and the light blue rectangle denotes the constraints $r<0.038$ and $n_{s}=0.9743 \pm 0.0034$}. {Compared to the first and second panels, the results favor a smaller $\frac{\delta}{V_{0}}$; the third and fourth panels also show this preference. Compared to the first and third panels, the results favor a smaller $n$; the second and fourth panels also show this preference. Compared to the cases $n=\frac{1}{2}$ and $n=\frac{1}{3}$, it can be seen from Figure~\ref{Fig3} that an increase in $n$ causes $n_{s}$  to shift to lower values and $r$ to shift to higher values. As is shown in all panels, an increase in $N$ leads to $n_{s}$ increasing and $r$ decreasing and an increase in $C$ leads to $n_{s}$ decreasing. In comparison with the observational constraints from P-ACT-LB-BK18, the results favor a larger $C$ and a smaller $n$. It is evident that an increasing $n$ or a decreasing $C$ could cause a mismatch with the observations. Nevertheless, for both cases $n=\frac{1}{2}$ and $n=\frac{1}{3}$, a suitable choice of $N$ and $C$ remains favored by the observational data.} Notably, the power-law potential~(\ref{Vp}) in standard slow-roll inflation is ruled out by the Planck 2018 results~\cite{Planck2020a} and is disfavored by ACT DR6~\cite{Calabrese2025}. However, slow-roll inflation is realized within RHDE and is supported by ACT DR6.

\begin{figure}[h]
%\begin{center}
\includegraphics[width=0.55\textwidth]{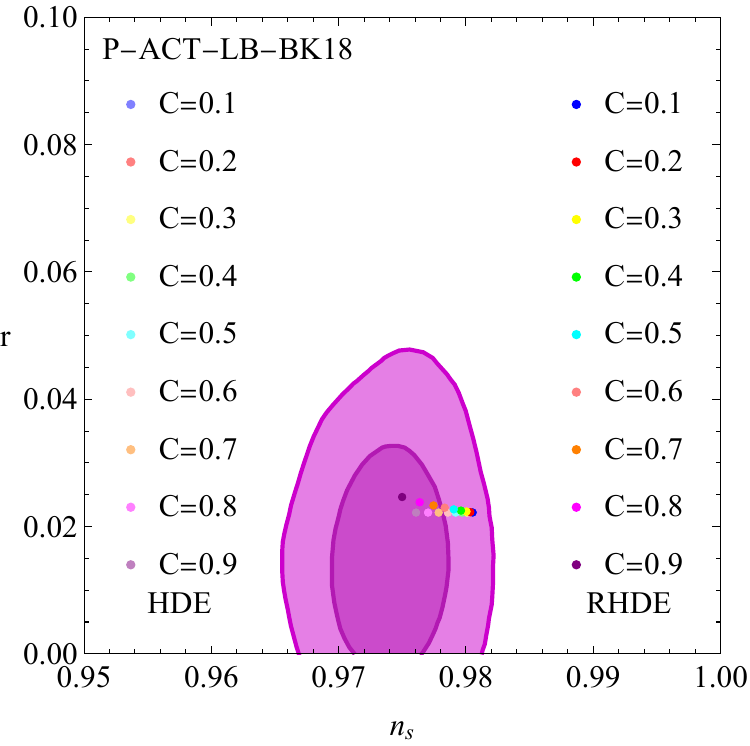}
\caption{\label{Fig3C} Predictions the power-law potential ~(\ref{Vp}) for slow-roll inflation with $N=60$ within HDE and RHDE in the $r$-$n_{s}$ plane. Light-colored points denote the results obtained from HDE, while dark-colored points correspond to RHDE results with $\frac{\delta}{V_{0}}=10^{-3}$.}
%\end{center}
\end{figure}

In Figure~\ref{Fig3C}, we compare the results for HDE and RHDE in the $r$-$n_{s}$ plane, for which we take $N=60$; light-colored points denote the results obtained from HDE, while dark-colored points correspond to RHDE results with $\frac{\delta}{V_{0}}=10^{-3}$. This figure shows that with the increase in $C$, the difference between RHDE and HDE becomes obvious, and RHDE has a larger $r$ and a smaller $n_{s}$.

\begin{figure}[h]
%\begin{center}
\includegraphics[width=0.6\textwidth]{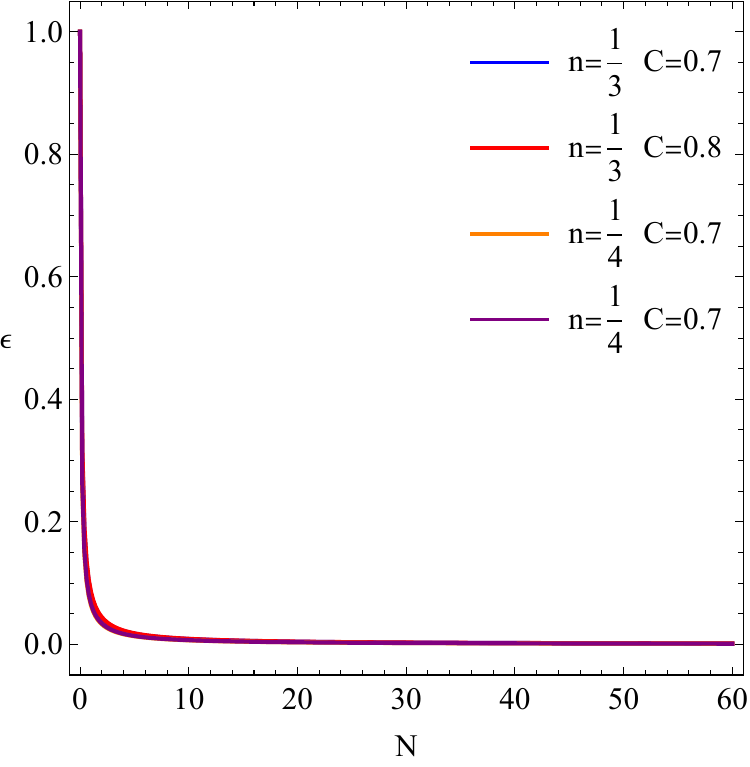}
\caption{\label{Fig3D} Slow-roll parameter $\epsilon$ versus $N$ with $\frac{\delta}{V_{0}}=10^{-3}$. Here, $N$ represents the remaining e-folds.}
%\end{center}
\end{figure}

To examine whether inflation ends naturally within the RHDE framework, we plot the evolutionary curves for the slow-roll parameter $\epsilon$ as a function of $N$, as shown in Figure~\ref{Fig3D}. Here, $N$ denotes the remaining e-folds, $N=60$ corresponds to the onset of the slow-roll regime, while $N=0$ denotes the end of inflation. With the decrease in $N$, $\epsilon$ increases and becomes equal to $1$ at $N=0$, indicating that the slow-roll approximation gradually breaks down and inflation terminates naturally. This behavior is consistent with the standard slow-roll inflation scenario and confirms that the RHDE framework does not prevent a graceful exit from inflation.

After inflation ended, the universe evolved into the hot expansion era, successively undergoing the radiation-dominated, matter-dominated, and dark energy-dominated epochs. To describe the evolution of the universe after inflation, dynamical analysis is proposed and widely used as a method for studying this process~\cite{Bahamonde2018, Huang2019a, Huang2021, Wu2010, Wu2008, Wu2007, Dutta2016, Dutta2017, Dutta2019, Huang2015, Chen2009, Chatterjee2025}. For RHDE, statefinder diagnostic~\cite{Sharma2020} and phase space analysis~\cite{Das2025} have been successfully used to analyze the evolution of the universe after inflation.

\subsection{Inflationary Attractor}

{It is well known that inflationary models with canonical scalar fields exhibit cosmological attractor behavior~\cite{Remmen2013}. This behavior plays a crucial role in viable scenarios by ensuring theoretical consistency across different initial conditions. The attractor behavior of the solution can be studied using both analytical and numerical methods. In this subsection, we analyze the attractor behavior within RHDE and examine whether our analytical solution acts as an attractor.}

\begin{figure}[h]
\begin{center}
\includegraphics[width=0.49\textwidth]{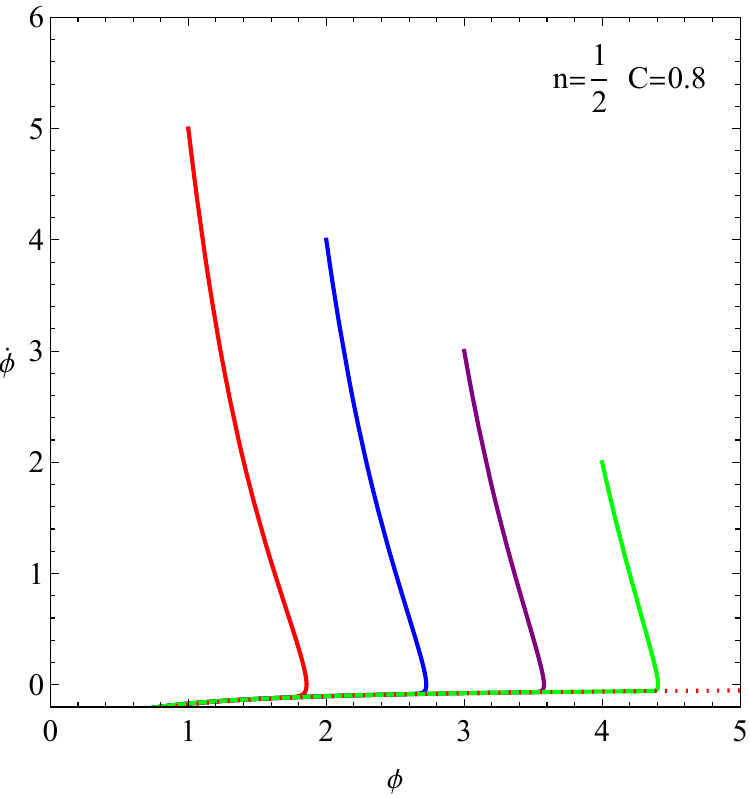}
\includegraphics[width=0.49\textwidth]{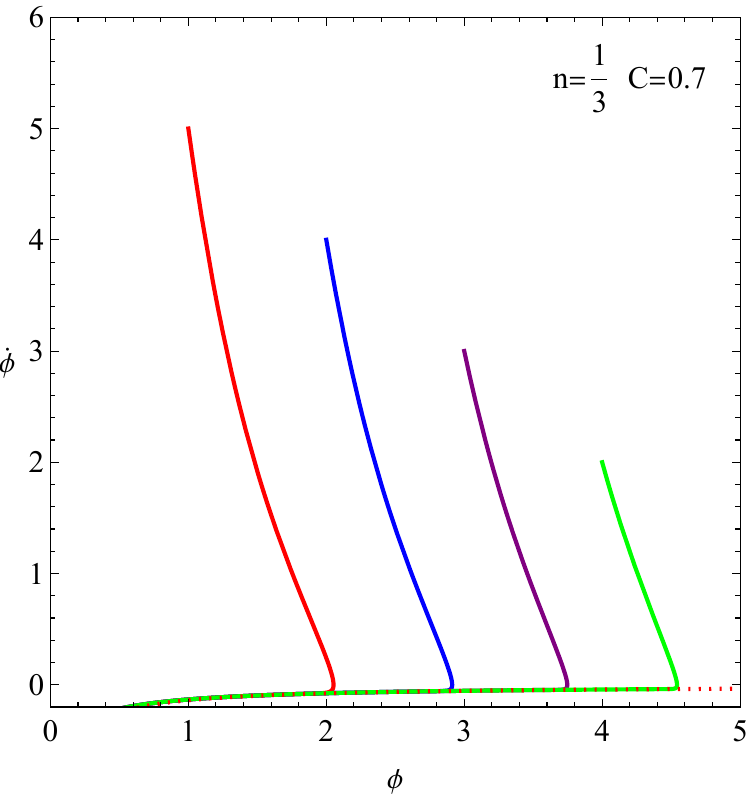}
\caption{\label{Fig4} Phase space diagram for the power-law potential ~(\ref{Vp}) within RHDE. The red dotted lines represent the analytical solution of the attractor, while the solid lines denote the evolutionary curves of the numerical solutions with different initial conditions.}
\end{center}
\end{figure}

Using Equations (\ref{H20}) and (\ref{phi20}), we obtain the analytical solutions for the slow-roll inflation within RHDE as follows\vspace{-10pt}
\beq
\dot{\phi}=-\frac{n V_{0}\phi^{n-1}}{3}\Bigg[\frac{6\kappa^{2}(1-C^{2})}{\kappa^{4}V_{0}\phi^{n}-24\pi^{2}\delta+\sqrt{576\pi^{4}\delta^{2}+48\kappa^{4}(1-2C^{2})\pi^{2}\delta V_{0}\phi^{n}+\kappa^{8}V_{0}^{2}\phi^{2n}}}\Bigg]^{\frac{1}{2}}.
\eeq
With different initial conditions, we numerically solve Equations (\ref{H2}) and (\ref{phi2}) and present the resulting phase space diagram in Figure~\ref{Fig4}, where the dotted lines represent the corresponding analytical solutions. These figures demonstrate that, regardless of the initial conditions, the kinetic energy rapidly decays to a negligible value, thereby driving the inflation field into a potential-dominated slow-roll state. The evolutionary trajectories exhibit strong convergence and ultimately approach the analytical solution indicated by the red-dotted lines. Consequently, within the framework of RHDE, the analytical solution serves as \mbox{an attractor.}

\subsection{Primordial Power Spectrum}

\begin{figure}[h]
%%begin{center}
\includegraphics[width=0.6\textwidth]{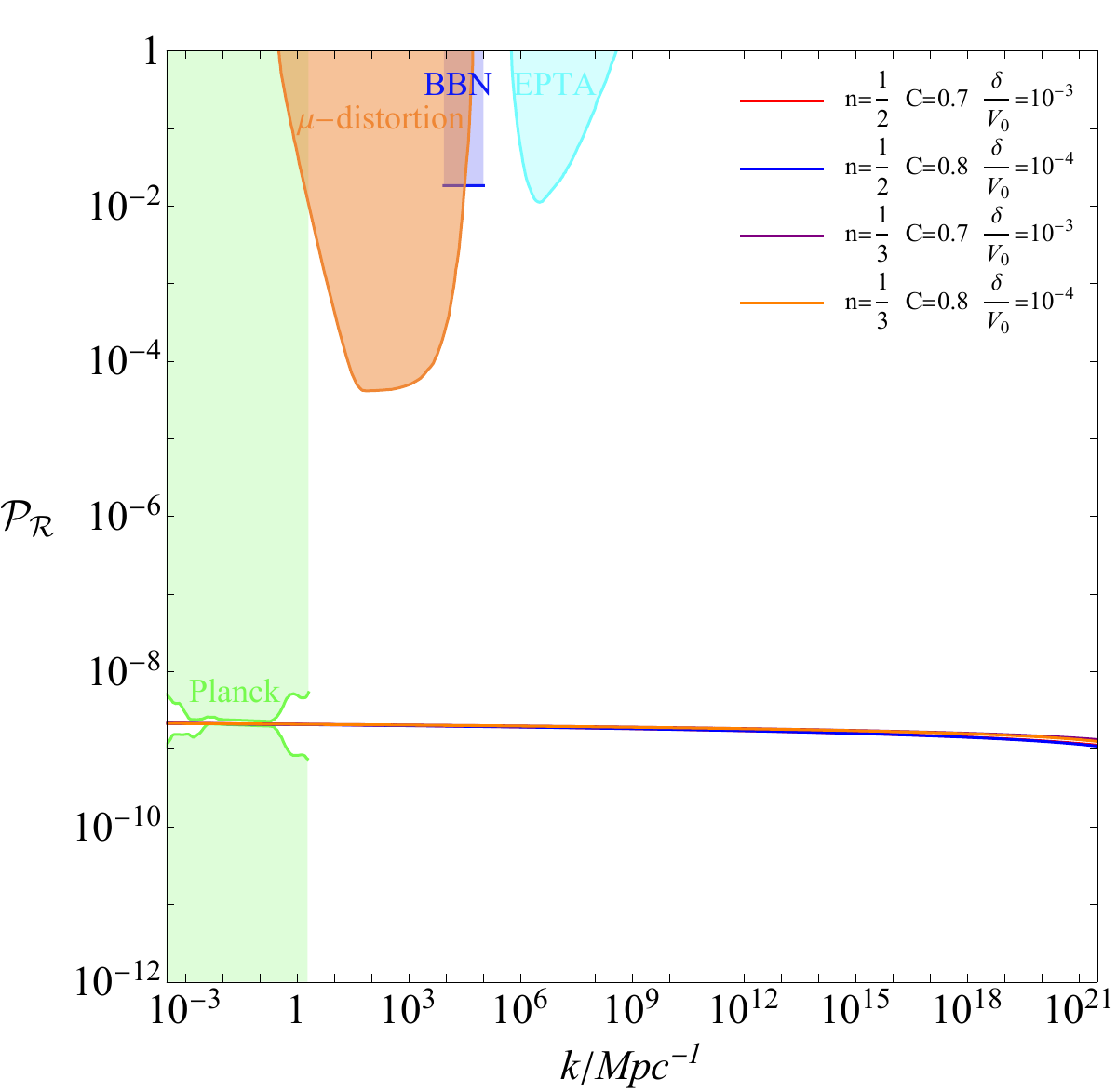}
\caption{\label{Fig5} Results for the primordial curvature perturbation power spectrum $\mathcal{P}_{\mathcal{R}}(k)$ within RHDE.}
%\end{center}
\end{figure}

{In the first subsection, we analyze the slow-roll inflation within RHDE and adopt the CMB pivot scale $k_{CMB} = 0.05Mpc^{-1}$. In this subsection, we present the full shape of the curvature power spectrum on all scales. To achieve this goal, under the slow-roll condition, we numerically solve the background Equations (\ref{H2}) and (\ref{phi2}) together with the Mukhanov–Sasaki equation
\beq
v_{k}''+\big( k^{2}-\frac{z''}{z} \big)v_{k}=0,
\eeq
where a prime $'$ denotes the derivative with respect to conformal time $\eta$, and $z=\frac{a \dot{\phi}}{H}$. Then, the power spectrum of primordial curvature perturbations $\mathcal{P}_{\mathcal{R}}(k)$ is given by
\beq
\mathcal{P}_{\mathcal{R}}(k)=\frac{k^{3}}{2\pi^{2}}\left| \frac{v_{k}}{z} \right|^{2}.
\eeq

Adopting the initial conditions listed in Table~\ref{Tab1}, which are chosen according to the results in the first subsection, we numerically solve these equations to obtain the primordial power spectrum. Our numerical results are listed in Table~\ref{Tab1} and shown in Figure~\ref{Fig5}, where they are superimposed on the observational data from CMB ~\cite{Planck2020}, $\mu$-distortion of CMB ~\cite{Fixsen1996}, big-bang nucleosynthesis (BBN ~\cite{Inomata2016}), and the European Pulsar Timing Array (EPTA ~\cite{Inomata2019}). It can be seen that our numerical results for the primordial curvature power spectra fall well within the bounds set by current observations. This confirms that the RHDE model, with the considered parameter space, is fully compatible with current multi-probe observational data. Therefore, the inflationary predictions derived from the RHDE framework are observationally viable and warrant further investigation in light of future high-precision measurements.}

\begin{table}[h]
\centering
\caption{\label{Tab1} Model parameters and the numerical results for primordial power spectrum within RHDE.}
 \begin{tabular}{|c|c|c|c|c|c|c|c|c|c|}
  \hline
  \hline
  \boldmath{$Model$} & \boldmath{$n$} & \boldmath{$C$} & \boldmath{$\delta$} & \boldmath{$V_{0}$} & \boldmath{$\frac{\delta}{V_{0}}$} & \boldmath{$N_{*}$} & \boldmath{$n_{s}$} & \boldmath{$r$} &\boldmath{ $\mathcal{P}_{\mathcal{R}}(k_{*})$}\\
  \hline
  $A$ & $\frac{1}{2}$ & $0.7$ & $1.28\times 10^{-10}$ & $1.28\times 10^{-7}$ & $10^{-3}$ & $59.53$ & $0.9748$ & $0.0342$ & $2.108\times 10^{-9}$\\
  \hline
  $B$ & $\frac{1}{2}$ & $0.8$ & $9.60\times 10^{-12}$ & $9.60\times 10^{-8}$ & $10^{-4}$ & $59.35$ & $0.9735$ & $0.0338$ & $2.103\times 10^{-9}$\\
  \hline
  $C$ & $\frac{1}{3}$ & $0.7$ & $4.18\times 10^{-10}$ & $4.18\times 10^{-7}$ & $10^{-3}$ & $59.77$ & $0.9773$ & $0.0234$ & $2.108\times 10^{-9}$\\
  \hline
  $D$ & $\frac{1}{3}$ & $0.8$ & $3.00\times 10^{-11}$ & $3.00\times 10^{-7}$ & $10^{-4}$ & $57.27$ & $0.9762$ & $0.0231$ & $2.105\times 10^{-9}$\\
  \hline
  \hline
  \end{tabular}
\end{table}

\section{Conclusions}

Based on the Rényi entropy, Rényi holographic dark energy has been proposed to explain the current accelerated phase of the universe. In this paper, we analyze holographic inflation and slow-roll inflation within the framework of RHDE. For the holographic inflation, we find it is ruled out by ACT DR6, since for $r < 0.06$, the predicted $n_{s}$ exceeds the observationally allowed range. For slow-roll inflation with the power-law potential $V_{0}\phi^{n}$, {we find that the observational constraints from P-ACT-LB-BK18 favor a larger $C$ and a smaller $n$, while for both cases $n=\frac{1}{2}$ and $n=\frac{1}{3}$, a suitable choice of $N$ and $C$ is favored by the observational data.

Then, we analyze the inflationary attractor within RHDE. By solving the background dynamical equation and plotting the evolutionary curves in $\dot{\phi}-\phi$ plane, we find that the evolutionary trajectories exhibit strong convergence and ultimately approach the analytical solution. This indicates that the analytical solution acts as an attractor. Additionally, we adopt the value of model parameters constrained by the observational data and numerically solve the background equations together with the Mukhanov–Sasaki equation, we plot the curves for the primordial curvature power spectrum and find our numerical results fall well within the bounds set by the observational data from CMB, $\mu$-distortion of CMB, big-bang nucleosynthesis, and the European Pulsar Timing Array. All results demonstrate that within the RHDE framework, slow-roll inflation is observationally viable, while holographic inflation is not supported by current observational data.

\begin{acknowledgments}

This work was supported by the National Natural Science Foundation of China under Grant Nos. 12405081, 12265019, and 11865018.

\end{acknowledgments}

\end{document}